# Towards Automated COVID-19 Presence and Severity Classification

Dominik MUELLER[a,b], Silvan MERTES[a], Niklas SCHROETER[a],
Fabio HELLMANN[a], Miriam ELIA[a], Bernhard BAUER[a], Wolfgang REIF[a],
Elisabeth ANDRÉ[a] and Frank KRAMER[a]
[a]*Faculty of Applied Computer Science, University of Augsburg, Germany*
[b]*Institute for Digital Medicine, University Hospital Augsburg, Germany*

**Abstract.** COVID-19 presence classification and severity prediction via (3D) thorax computed tomography scans have become important tasks in recent times. Especially for capacity planning of intensive care units, predicting the future severity of a COVID-19 patient is crucial. The presented approach follows state-of-the-art techniques to aid medical professionals in these situations. It comprises an ensemble learning strategy via 5-fold cross-validation that includes transfer learning and combines pre-trained 3D-versions of ResNet34 and DenseNet121 for COVID19 classification and severity prediction respectively. Further, domain-specific preprocessing was applied to optimize model performance. In addition, medical information like the infection-lung-ratio, patient age, and sex were included. The presented model achieves an AUC of 79.0% to predict COVID-19 severity, and 83.7% AUC to classify the presence of an infection, which is comparable with other currently popular methods. This approach is implemented using the AUCMEDI framework and relies on well-known network architectures to ensure robustness and reproducibility.

## 1. Introduction

In response to the rapid spread of the SARS-CoV-2 virus at the beginning of the year 2020, many scientists quickly reacted and developed various approaches based on deep learning to contribute to the efforts against COVID-19. Especially the severity assessment of patients is essential for treatment decisions and disease course monitoring. However, the course of the SARS-CoV-2 virus pandemic showed that one of the most critical factors for COVID-19 treatment is the capacity of intensive care units (ICU) at hospitals (1; 2). Through the rapid but inconsistent development of infection severity, predicting the future severity of a patient within a month (prognosis) for capacity planning of ICUs is challenging but crucial (1; 2). This paper presents an approach to assess COVID-19 severity through the classification of 3D thorax computed tomography (CT) scans. It relies on the in-house AUCMEDI framework, which provides easy access to pretrained models, ready for medical transfer learning. This approach enables swift adoption since the heavy lifting is moved to the framework. Our method achieves promising results, and we encourage other researchers to build on the presented baseline.

2. **Methods**

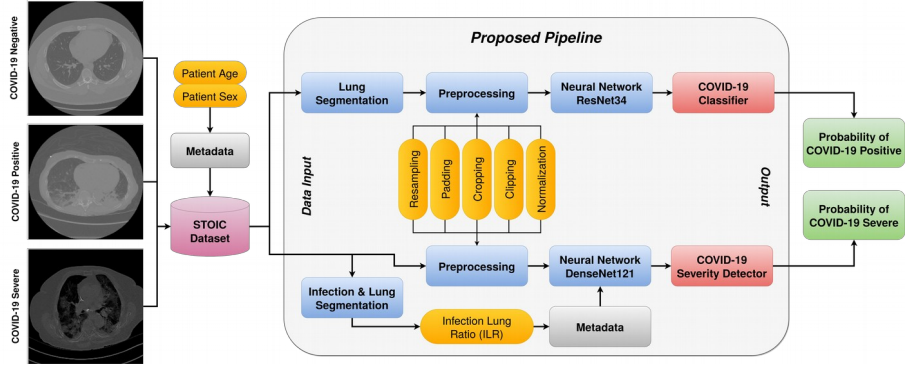

Figure 1. Flowchart diagram of the proposed pipeline.

To set up a modern and effective medical image classification pipeline (Figure 1), the in-house framework AUCMEDI was utilized (3), a software package that offers a library as a high-level API for the standardized construction of modern medical image classification pipelines. As illustrated in Figure 1, our model takes thorax CT scans as input, in addition to the patient's age, sex, and *Infection-Lung Ratio* (ILR), which is computed a priori. The outputs of the model are estimated probabilities for COVID-19 presence and a severe outcome within a month. An ensemble approach was applied to assess the presence and severity of COVID-19 based on CT scans of the thorax, in which independent models were used for the individual prediction of presence with our 'COVID-19 Classifier' and severity with our 'COVID-19 Severity Detector'. For model training and internal performance evaluation, the dataset from the STOIC (Study of Thoracic CT in COVID-19) project was used (4).

*1.1. Pre-processing and Image Augmentation*

The following image augmentation methods were applied for the training process: rotation, flipping, scaling, gamma modification, and elastic deformations. In contrast to the COVID-19 Severity Detector, a segmentation of the lung is applied for the COVID19 Classifier workflow before pre-processing. Pixels outside of the segmented lung region are excluded and the volume is cropped to a minimal shape around the lung. For the pre-processing of both models, the CT volume is resampled to a voxel spacing of 1.48x1.48x2.10 mm and clipped to the range −1024 Hounsfield Units (HU) to +100 HU to exclude irrelevant tissue types as well as to reduce complexity. Subsequently, the pixel intensities of the volume were standardized to a grayscale range. Samples that might exceed the accepted input image size of 148x224x224 pixels are center-cropped or padded if undersized to match the required image size. For training, random cropping was applied instead of center-cropping. As the last pre-processing step, another normalization is applied via the Z-Score normalization approach based on the mean and standard deviation computed on the ImageNet dataset (5). The severity of lung diseases is commonly quantified in radiology by the percentage of lung area which is affected. This ratio between affected and healthy lung tissue provides intuitive insights into the current state and severity of the disease. For pneumonia as well as COVID-19, a similar approach can be applied. The ILR describes the ratio between infected and healthy tissue in the lung measured in pixels: $ILR = |Infection|/(|Infection|+|Lung|)$. Several studies (6; 7; 8) proved that the ILR assessed on CT scans has a high correlation to severity measured by labor biomarkers as well as the survival rate. Thus, the computed ILR was integrated as metadata in the COVID-19 Severity Detector, utilizing the pipeline by (9).

*1.2. Neural Network Models and Ensemble Learning Strategy*

For the COVID-19 Classifier, a 3D version of the ResNet34 (10) architecture is used, whereas the COVID-19 Severity Detector is based on a 3D version of the DenseNet121 (11) architecture. The COVID-19 Severity Detector classification head was modified to additionally take metadata into account. The metadata consists of three parts: Patient age, sex, and the ILR of each sample. Regarding the training process of both models, transfer learning was applied to the classification head, and a fine-tuning strategy on all layers. The transfer learning was conducted for 10 epochs, using the Adam optimizer with an initial learning rate of $1 \times 10^{-4}$ and a batch size of 4 for the DenseNet, and 8 for the ResNet. The fine-tuning run for a maximum of 240 epochs, using a dynamic learning rate starting from $1 \times 10^{-5}$ to a maximum decrease of $1 \times 10^{-7}$ (decreasing factor of 0.1 after 8 epochs without improvement on the monitored validation loss). Furthermore, an early stopping technique was utilized, stopping after 36 epochs without improvement. As a loss function, the sum of the F1-score and the weighted Focal loss from Lin et al. (12) was utilized. For inference, the model with the best validation loss is used. The COVID-19 Classifier predicts presence and severity, resulting in the following three classes: Negative, Positive, and Severe (subset of Positive).

A 5-fold cross-validation was applied on both models as a Bagging approach for ensemble learning. For the performance evaluation, the combined cross-validation folds for training were split into a training (70%) and validation subset (10%), whereas the remaining cross-validation fold (20%) was utilized for testing.

The prediction of the final COVID-19 Severity Detector pipeline comprises the mean-averaged sum of all five predictions from the five models of the cross-validation. This approach not only allows for more efficient usage of the available data but also increases the reliability of the prediction. However, for the COVID-19 Classifier, only the model with the best-monitored validation loss was used for predicting the COVID-19 Positive class in the final pipeline. Internal experiments revealed that this allowed achieving the best testing performance for COVID-19 presence prediction compared to utilizing a similar pooling-based prediction strategy like the COVID-19 Severity Detector.

2. **Results**

By utilizing cross-validation for evaluation, we were able to compute performance metrics on all samples (Table 1). For the COVID-19 Classifier, a multi-class evaluation was performed in which the class with the highest predicted probability was used as the outcome of the evaluation. As metrics, the Accuracy, F1-Score, and AUC were computed.

| Classifier | | | |
|---|---|---|---|
| Classes | Acc. | F1 | AUC |
| Negative | 0.787 | 0.782 | 0.837 |
| Positive | 0.701 | 0.629 | 0.766 |
| Severe | 0.782 | 0.424 | 0.788 |
| Classifier and Severity Detector | | | |
| Method | Probability | | AUC |
| Classifier | infection | | 0.837 |
| Detector | severe outcome | | 0.790 |

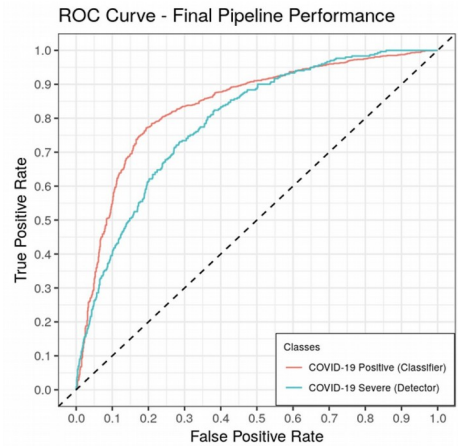

Table 1. Computed performance results of the COVID-19 Classifier and Severity Detector.

Figure 2. Results of the performance evaluation for COVID-19 infection and severe outcome prediction.

For the final performance assessment, the predicted probabilities for the COVID-19 infection and severe outcomes were evaluated by the AUC metric. The COVID-19 infection prediction achieved an AUC of 83.7%, whereas the severe COVID-19 outcome prediction had an AUC of 79.0% (Figure 2). In contrast, the COVID-19 Classifier obtained an AUC of 78.8% for severity outcome prediction.

3. **Discussion**

The proposed pipeline demonstrated robust classification performance for predicting COVID-19 infection and severe outcomes within a month. This is why a separate model for severity prediction with the COVID-19 Severity Detector was implemented instead of utilizing the inferior COVID-19 Severe class prediction of the COVID-19 Classifier. Overall, accurately predicting the severe outcome within a month is challenging. Analyzing the prediction results of the COVID-19 Classifier revealed that especially the differentiation between COVID-19 Positive and COVID-19 Severe patients is a hard and complex task.

4. **Conclusions**

In this study, a powerful pipeline for COVID-19 infection and severe outcome prediction was implemented, utilizing the AUCMEDI framework combined with integrated ensemble, transfer, and deep learning techniques. The pipeline consisted of a multi-model workflow for individual prediction of the outcome variables to ensure utilization of the best-suited model design. By utilizing a standardized framework, our approach is easily reproducible which guarantees a fast adoption for other projects. In future work, additional validation and integration in a clinical study are needed to identify actual medical gain in a clinical workflow.